\documentclass[english,10pt,letterpaper,twocolumn,superscriptaddress,floatfix,allowfontchangeintitle,accepted=2025-02-03]{quantumarticle}
\pdfoutput=1
\usepackage[utf8]{inputenc}
\usepackage[T1]{fontenc}
\usepackage{amsmath}
\usepackage{amssymb}
\usepackage{amsfonts}
\usepackage{bm}
\usepackage{bbm}
\usepackage{epsfig}
\usepackage{grffile}
\usepackage{graphics}
\usepackage{enumitem}

\usepackage[numbers,sort&compress]{natbib}
\usepackage{scalefnt}

\usepackage[usenames,dvipsnames]{xcolor}
\definecolor{dblue}{rgb}{0,0.1,.6}

\newcommand{\Emph}[1]  {{\emph{#1}}}

\usepackage[colorlinks=true,citecolor=dblue,linkcolor=dblue,urlcolor=dblue]{hyperref}
\usepackage[all]{hypcap}

\newcommand{\ud}{\mathrm{d}}
\newcommand{\id}{\mathbbm{1}}
\newcommand{\bra}{\langle}
\newcommand{\ket}{\rangle}
\newcommand{\mc}[1]{\mathcal{#1}}
\renewcommand{\vec}[1]{{\boldsymbol{#1}}}
\newcommand{\mri}{\mathrm{i}\mkern1mu}

\renewcommand{\O}{\mc{O}}

\newcommand{\hs}{{\hat{\sigma}}}
\newcommand{\hH}{\hat{H}}
\newcommand{\hh}{\hat{h}}
\newcommand{\hR}{\hat{R}}
\newcommand{\hS}{\hat{S}}
\newcommand{\hvS}{\hat{\vec{S}}}
\newcommand{\hU}{\hat{U}}

\newcommand{\gs}{\text{gs}}

\newcommand{\lin}[2]{\xrightarrow{(#1,#2)}}

\newcommand{\duke} {Department of Physics, Duke University, Durham, North Carolina 27708, USA}
\newcommand{\dqc}  {Duke Quantum Center, Duke University, Durham, North Carolina 27701, USA}
\newcommand{\tz}   {Tensor Center, Auf dem Dresch 15, 52152 Simmerath, Germany}

\newcommand{\Title} {Convergence and quantum advantage of Trotterized MERA for strongly-correlated systems}

\begin{document}

\title{\scalefont{0.97}\Title}
\author{Qiang Miao}
\affiliation{\dqc}
\author{Thomas Barthel}
\affiliation{\dqc}
\affiliation{\duke}
\affiliation{\tz}

\date{September 9, 2024}

\begin{abstract}
Strongly-correlated quantum many-body systems are difficult to study and simulate classically.
We recently proposed a variational quantum eigensolver (VQE) based on the multiscale entanglement renormalization ansatz (MERA) with tensors constrained to certain Trotter circuits. Here, we determine the scaling of computation costs for various critical spin chains which substantiates a polynomial quantum advantage in comparison to classical MERA simulations based on exact energy gradients or variational Monte Carlo. Algorithmic phase diagrams suggest an even greater separation for higher-dimensional systems. Hence, the Trotterized MERA VQE is a promising route for the efficient investigation of strongly-correlated quantum many-body systems on quantum computers.
Furthermore, we show how the convergence can be substantially improved by building up the MERA layer by layer in the initialization stage and by scanning through the phase diagram during optimization.  For the Trotter circuits being composed of single-qubit and two-qubit rotations, it is experimentally advantageous to have small rotation angles. We find that the average angle amplitude can be reduced considerably with negligible effect on the energy accuracy. Benchmark simulations suggest that the structure of the Trotter circuits for the TMERA tensors is not decisive; in particular, brick-wall circuits and parallel random-pair circuits yield very similar energy accuracies.
\end{abstract}

\maketitle
\hypersetup{ pdftitle = {} }

\section{Introduction}\label{sec:intro}
Due to the quantum advantage \cite{Arute2019-574,Zhong2020-370} for certain tasks, quantum computers are expected to lead to a revolution in information technology. While there has been enormous theoretical and experimental progress, quantum computation currently still faces a number of challenges, including the limited number of qubits, inevitable noise, and insufficient fidelity \cite{Bharti2022-94,Stilck2021-17}. The simulation of quantum matter is less demanding and, hence, a promising field of application for the current era of noisy intermediate-scale quantum (NISQ) devices \cite{Wecker2015-92,Liu2019-1,Smith2019-5,Motta2020-16,Cade2020-102,Miao2021_08,Barratt2021-7,FossFeig2021-3,Niu2022-3,Chertkov2022-18}. Compared to classical techniques, quantum simulation methods can have substantially lower time complexity and may, hence, lead to new insights on quantum many-body physics in the near future.

In Ref.~\cite{Miao2021_08} we proposed a resource-efficient and noise-resilient \cite{Kim2017_11} variational quantum eigensolver (VQE) for the simulation of strongly-correlated quantum many-body systems. This VQE is based on the multi-scale entanglement renormalization ansatz (MERA) \cite{Vidal-2005-12,Vidal2006} and gradient-based optimization. A strength of tensor network state (TNS) \cite{Baxter1968-9,White1992-11,Niggemann1997-104,Verstraete2004-7,Vidal-2005-12,Vidal2006,Schollwoeck2011-326,Orus2014-349} techniques such as MERA is that they are applicable for frustrated quantum magnets and strongly-correlated fermionic systems \cite{Barthel2009-80,Corboz2009-80,Kraus2009_04,Corboz2009_04,Pineda2009_05,Mortier2025-18}, where quantum Monte Carlo is hampered by the sign problem \cite{Takasu1986-75,Loh1990-41,Chandrasekharan1999-83,Troyer2005}. Nevertheless, a limiting factor for classical TNS simulations is the high computational cost, especially for $D\geq 2$ spatial dimensions. We aim to overcome this problem by combining MERA with the VQE approach. In order to allow for an efficient experimental implementation, the MERA (dis)entanglers and isometries are chosen as circuits of two-qubit gates. We refer to the latter as Trotter gates, because they get closer and closer to the identity when increasing the number of gates per MERA tensor \cite{Miao2021_08}. This construction leads to what we call Trotterized MERA (TMERA)\footnote{A similar structure, which is called brick-wall qMERA, was proposed simultaneously in Ref.~\cite{Haghshenas2022-12}. The expressiveness of quantum circuits was studied numerically for a 1D binary qMERA with $N=32$ sites. We focus on substantiating a quantum advantage by comparing the scaling of computation costs for quantum and classical MERA algorithms, using algorithmic phase diagrams and benchmark simulations for large 1D systems.}. Recently, the VQE approach has been applied to small quantum chemistry problems \cite{OMalley2016-6,Kandala2017-549,Colless2018-8,Hempel2018-8,Mcardle2019-5,Grimsley2019-10,Nam2020-6,Self2021-7,Guo2024-20}. We expect that the TMERA VQE will lead to similar progress for quantum many-body problems, particularly for strongly-correlated systems which cannot be easily simulated on classical computers.

In this paper, we try to establish a quantum advantage of the TMERA VQE over classical MERA algorithms and describe further improvements for its  practical implementation.
Sec.~\ref{sec:converge} describes and analyzes different schemes for initializing the TMERA optimization and driving it to convergence with the goal of avoiding local minima. Iteratively increasing the number of variational parameters can substantially improve the performance. One can also start from system parameters with a well-known low-entangled ground state and, during the optimization, then move on certain paths through the space of model parameters to obtain the actual ground state of interest.
In the central Secs.~\ref{sec:complexity} and \ref{sec:qAdvantage}, we analyze the quantum-computational complexity of TMERA VQE and compare to the classical computation costs of (unconstrained) full MERA (fMERA) as well as TMERA methods, taking into account the stochastic nature of quantum measurements and considering the classical methods employing exact energy gradients (EEG) \cite{Evenbly2009-79} as well as a variational Monte Carlo (VMC) scheme \cite{Ferris2012-85,Barthel2025-111}. For several critical one-dimensional spin-$s$ quantum magnets, we find a polynomial quantum advantage for the TMERA VQE over the classical MERA methods. In the considered examples, the quantum advantage increases with increasing spin quantum number $s$. We determine algorithmic phase diagrams that suggest considerably larger quantum advantages for systems in $D\geq 2$ spatial dimensions.
Small two-qubit rotation angles are desirable for implementations on present-day quantum devices. In Sec.~\ref{sec:anglePenalty}, we add an angle penalty term to the energy functional and find that the average angle can be reduced considerably without substantially affecting energy accuracies.
The initial TMERA proposal \cite{Miao2021_08} chose all TMERA tensors as brick-wall circuits (a.k.a.\ alternating layered ansatz) with generic two-qubit gates acting on nearest neighbors. In Sec.~\ref{sec:PRPC}, we replace these brick-wall circuits by parallel random-pair circuits (PRPC). While PRPC contain long-range two-qubit gates, they do not seem to outperform brick-wall circuits, as long as bond dimensions are relatively small. Section~\ref{sec:discuss} summarizes the findings and comments on the broader context.

\section{TMERA and tensor circuit structure}\label{sec:mera}
\begin{figure}[t]
	\includegraphics[width=\columnwidth]{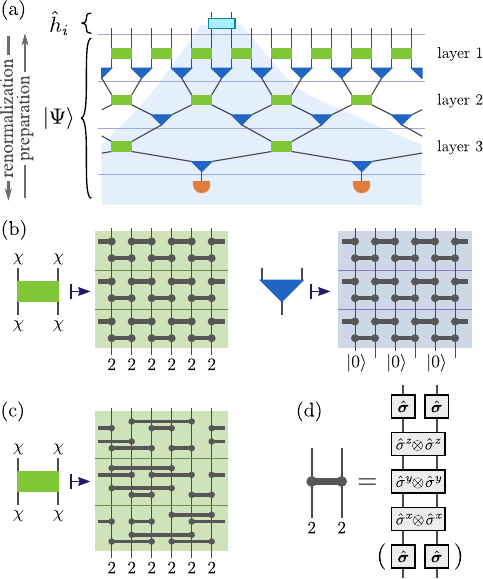}
	\caption{\label{fig:MERA} \textbf{Trotterized MERA and circuit structures}. (a) A binary MERA with $T=3$ layers. The shaded region indicates the causal cone of a two-site operator $\hh_i$. All tensors of a TMERA -- the unitary (dis)entanglers and the isometries -- are chosen as Trotter circuits. Panel (b) shows a corresponding brick-wall circuit consisting of nearest-neighbor two-qubit gates with $t=3$ Trotter steps, and bond dimension $\chi=2^3$, i.e., $q=3$ qubits per renormalized site. Panel (c) illustrates the alternative choice of a parallel random-pair circuit. (d) Each two-qubit gate can be realized by single-qubit and two-qubit rotations.}
\end{figure}

Let us consider a lattice system with $N$ sites, each associated with a single-site Hilbert space of dimension $d$. If a VQE completely replicates the physical system, the required number of qubits grows linearly in the system size $N$. This can be avoided by employing tensor network structures \cite{Huggins2019-4,Liu2019-1,FossFeig2021-3,Miao2021_08,Slattery2021_08,Niu2022-3,Chertkov2022-18}. We consider MERA $|\Psi\ket$ as shown in Fig.~\ref{fig:MERA} that are suitably adapted for VQE. The hierarchical structure of MERA is motivated by real-space renormalization group. In each layer $\tau$, degrees of freedom of the system are disentangled to some extent by unitaries before isometries reduce the total number of sites by a factor of $b$ in the transition to layer $\tau+1$; $b$ is called the branching ratio. After $T\lesssim \log_b(N)$ layers the procedure is stopped by projecting each site onto a reference state. Seen in reverse, this procedure prepares a quantum state for the physical system. Note that one can also work directly in the thermodynamic limit $N\to\infty$. For example, the tensor network in Fig.~\ref{fig:MERA}a can be interpreted as an elementary cell and repeated ad infinitum.

The expectation value $\bra\Psi|\hh_i|\Psi\ket$ of a local operator only depends on the MERA tensors in the causal cone of $\hh_i$. As a result, at most $\O(T)$ qubits are needed to evaluate energy expectation values and gradients. This is because only a small constant number of new qubits per layer is introduced in each layer transition $\tau\to\tau-1$ to realize the isometries. As the same number of qubits leaves the causal cone in each layer transition, mid-circuit resets \cite{Reed2010-96,Magnard2018-121,Egger2018-10,Schindler2011-332,Gaebler2021-104} can eliminate the $T$ dependence completely \cite{Miao2021_08}. The accuracy of MERA is controlled by the bond dimension $\chi$ of the tensors (equal to the Hilbert-space dimension of each renormalized site) and the number $T$ of layers. On the quantum computer, each renormalized site is represented by $q$ qubits such that $\chi=2^q$.

It is possible to implement this protocol for a full MERA on a quantum computer, but imposing an additional structure/constraint on the MERA tensors will greatly enhance the performance. Figure~\ref{fig:MERA} illustrates the idea of TMERA: A Trotter structure is imposed on every disentangler and isometry of the MERA. In Fig.~\ref{fig:MERA}b, each tensor is chosen as a brick-wall circuit with $t$ Trotter steps, which contain nearest-neighbor two-qubit unitary gates. As a result, the quantum circuits for the evaluation of energy expectation values and gradients will have depth $\O(tT)$. Figure~\ref{fig:MERA}c shows an alternative Trotter structure -- a parallel random-pair circuit (PRPC) consisting of arbitrary-range two-qubit gates. Per Trotter step, the PRPC contains the same number of gates as in the brick-wall circuit, but the qubit-pairs are chosen randomly. In both cases, the universal two-qubit gates can be implemented by a sequence of single-qubit rotations and CNOTs or, equivalently, single and two-qubit rotations \cite{Khaneja2000_10,Kraus2001-63,Zhang2003-67}; see Fig.~\ref{fig:MERA}d.

\section{Models}\label{sec:models}
To benchmark TMERA, we simulate various translation invariant models
\begin{equation}
	\hH=\sum_i \hh_i
\end{equation}
for one-dimensional quantum magnets, where $\hh_i$ are finite-range interaction terms. The models are all critical and have, hence, highly-entangled ground states with the long-range physics described by 1+1 dimensional conformal field theories (CFT) \cite{Belavin1984-241,Francesco1997}. We simulate directly in the thermodynamic limit ($N\rightarrow\infty$) with the groundstate energy per site denoted by $e^{\infty}_{\gs}=\lim_{N\to\infty}E/N$, and $c$ being the central charge of the CFT.

\Emph{Spin-1/2 models.}~--- The spin-1/2 XXZ chain \cite{Cloizeaux1966-7,Johnson1972-6,Mikeska2004}
\begin{equation}\label{eq:XXZ}
	\hH_{{1}/{2}} = \sum_i (\hS^x_i \hS^x_{i+1} + \hS^y_i \hS^y_{i+1} + \Delta \hS^z_i \hS^z_{i+1})
\end{equation}
is Bethe-ansatz integrable \cite{Bethe1931,Korepin1993}, where $\Delta$ is an anisotropy parameter. There is a critical spin-liquid phase when $|\Delta| \leq 1$, and the long-range physics can be explained by the sine-Gordon quantum field theory, which is a CFT with central charge $c = 1$. For $|\Delta| <1$, the groundstate energy density is \cite{Yang1966-150,Yang1966-150b}
\begin{equation*}
	e^\infty_\gs=\frac{\Delta}{4} - \int\ud x\, \frac{1- \Delta^2}{2\cosh(\pi x)\big[\cosh(2x\arccos{\Delta} ) - \Delta \big]}.
\end{equation*}
At the Berezinskii-Kosterlitz-Thouless (BKT) phase transition point $\Delta = 1$,  we recover the isotropic Heisenberg antiferromagnet (XXX chain) $\hH_{{1}/{2}} = \sum_i \hvS_i\cdot\hvS_{i+1}$ with $e^{\infty}_\gs = \frac{1}{4} - \ln{2}$ \cite{Hulthen1938}.

\Emph{Spin-1 models.}~--- The bilinear-biquadratic (BLBQ) spin-1 chain \cite{Uimin1970-12,Lai1974-15,Sutherland1975-12,Takhtajan1982-87,Babujian1982-90,Babujian1983-215,Laeuchli2006-74,Binder2020-102}
\begin{equation}\label{eq:BLBQ}
	\hH_{1} = \sum_i\big[\cos\vartheta (\hvS_i\cdot\hvS_{i+1}) + \sin\vartheta (\hvS_i\cdot\hvS_{i+1})^2\big]
\end{equation}
has several interesting quantum phases for the ground state. At the Takhtajan-Babujan (TB) point, $\vartheta = -\frac{\pi}{4}$, the model is Bethe-ansatz integrable and critical with central charge $c = \frac{3}{2}$ and groundstate energy density $e^\infty_\gs = -2\sqrt{2}$ \cite{Takhtajan1982-87,Babujian1982-90,Babujian1983-215}. As $\vartheta$ increases, we pass through the gapped Haldane phase, $-\pi/4 < \vartheta < \pi/4$, followed by a critical phase for $\pi/4 \leq \vartheta < \pi/2$. At the SU(3)-symmetric Uimin-Lai-Sutherland (ULS) point, $\vartheta = \frac{\pi}{4}$, the system can be solved by the nested Bethe ansatz \cite{Uimin1970-12,Lai1974-15,Sutherland1975-12}, is described by a CFT with central charge $c = 2$, and has groundstate energy density $e^\infty_\gs = -\frac{\sqrt{2}}{2}  (\ln{3} + \frac{\pi}{3\sqrt{3}} - 2)$ \cite{Uimin1970-12}.

\Emph{Spin-3/2 models.}~--- The bilinear-biquadratic-bicubic (BLBQBC) spin-3/2 chain
\begin{equation}\label{eq:BLBQBC}\textstyle
	\hH_{{3}/{2}} = \sum_i \Big[-\frac{1}{16} \hvS_i\cdot\hvS_{i+1} + \frac{1}{54} (\hvS_i\cdot\hvS_{i+1})^2 + \frac{1}{27} (\hvS_i\cdot\hvS_{i+1})^3\Big]
\end{equation}
is also Bethe-ansatz integrable \cite{Takhtajan1982-87,Babujian1982-90,Babujian1983-215}. Similar to the BLBQ spin-1 model at the TB point, it is an integrable analog of the isotropic spin-1/2 Heisenberg antiferromagnet for spin-3/2. It has a central charge $c = \frac{9}{5}$, and a groundstate energy density of $e^{\infty}_\gs = -\ln{2} - \frac{1}{8}$ \cite{Alcaraz1988-21}.

Lastly, we consider the spin-$3/2$ XXX chain,
\begin{equation}\label{eq:XXX3d2}
	\hH_{{3}/{2}}'=\sum_i \hvS_i\cdot\hvS_{i+1},
\end{equation}
which has central charge $c = 1$ \cite{Schulz1986-34,Affleck1987-36,Hallberg1996-76} and  groundstate energy density $e^{\infty}_\gs \approx -2.828\ 33$ \cite{Hallberg1996-76,Lou2002-65,Ramos2014-89}. While this half-integer spin model has similar critical properties like the spin-$1/2$ XXX chain \cite{Haldane1983-93,Haldane1983-50,Schulz1986-33,Schulz1986-34,Alcaraz1992-46,Hallberg1996-76}, it is non-integrable and features higher entanglement due to the larger spin quantum number \cite{Xavier2010-81,Dalmonte2012-85}.

\section{\texorpdfstring{\!\!\!Initialization and convergence schemes}{Initialization and convergence schemes}}\label{sec:converge}
To find a global minimum in a nonlinear optimization problem is generally non-trivial. The variational energy minimization for a many-body system can easily converge to local minima. We try the following approaches to address this problem for the TMERA VQE:
\begin{itemize}
  \item
  \emph{Direct TMERA:}
  One may simply optimize the TMERA without modifications, starting with a random initialization.
  \item 
  \emph{TTTN$\,\mapsto$TMERA:}
  It can be beneficial to begin with a reduced set of parameters and then gradually increase the complexity of the variational ansatz. Specifically, we may first remove all disentanglers (set them to $\mathbbm{1}$) such that the TMERA becomes a Trotterized tree tensor network (TTTN) \cite{Fannes1992-66,Otsuka1996-53,Shi2006-74,Murg2010-82,Tagliacozzo2009-80}. The converged TTTN, is then used to initialize the TMERA.
  \item
  \emph{Built-up TMERA:}
  A similar idea is to build up the TMERA tensor network layer by layer, i.e., iteratively increase the number of parameters during the optimization\footnote{One can also increase the number of parameters by iteratively increasing the number of Trotter steps \cite{Haghshenas2022-12}.}. This is motivated by the decay of average energy-gradient amplitudes with $\tau$ \cite{Barthel2023_03,Miao2024-109}.
  \item
  \emph{Scanning:}
  A powerful method is to start the optimization at a point in the phase diagram of the model, where we know the ground state very well or know, for example, that it is low-entangled such that it can be easily determined. In a ``scanning'' procedure, one then moves on suitable paths through the parameter space of the Hamiltonian during the optimization, until obtaining the actual ground state of interest.
\end{itemize}
\begin{figure}[t]
	\includegraphics[width=\columnwidth]{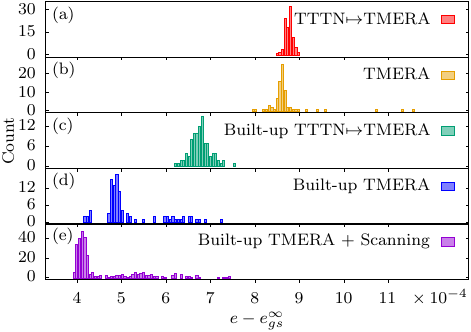}
	\caption{\label{fig:dist} \textbf{Convergence schemes.} The figure shows histograms of converged energy densities $e$ for the spin-1/2 XXZ model~\eqref{eq:XXZ} at $\Delta = 1$ using homogeneous modified binary TMERA with $T = 6$ layers, $t = 8$ Trotter steps, and bond dimension $\chi = 8$ ($q = 3$). From top to bottom, the convergence methods are: (a) optimizing a TTTN with $T = 6$, and then switching to TMERA; (b) directly optimizing a TMERA with $T = 6$; (c) building up a TTTN layer by layer and then switching to TMERA; (d) building up a TMERA layer by layer; and (e) building up a TMERA layer by layer at $\Delta = 2$ and then scanning in the range $\Delta\in[0,2]$.}
\end{figure}
One can also combine these ideas, e.g., by first building up a TTTN and then switching to TMERA.

In every scheme, we initialize all isometric tensors randomly and all disentanglers as identities at the beginning of the optimization procedure. Note that this could, in principle, lead to the barren-plateau phenomenon of exponentially small initial gradients \cite{McClean2018-9,Cerezo2021-12}. In Refs.~\cite{Barthel2023_03,Miao2024-109}, we show that TMERA are not hampered by this problem. For tensors of the first layer ($\tau=1$), the average gradient of random tree tensor networks and MERA decreases polynomially with increasing bond dimension $\chi$. For higher layers, it decreases as $(b\eta)^\tau$, where $\eta$ is the second largest eigenvalue of a (doubled) layer transition channel, which is also polynomial in $\chi$ \cite{Barthel2023_03,Miao2024-109}.

To test the performance of the different convergence methods, we run each for a few hundred times, starting from different random initializations. 
Figure~\ref{fig:dist} shows results for the spin-1/2 XXZ chain with homogeneous modified binary TMERA using a Riemannian L-BFGS optimization \cite{Huang2015-25,Hauru2021-10,Luchnikov2021-23,Miao2021_08}, i.e., a quasi-Newton method. The resulting TMERA energy densities $e$ are compared to the exact value $e^\infty_\gs$. While the $e$ distributions for TTTN$\,\mapsto$TMERA and direct TMERA are narrow, without building up, their average and lowest $e$ are considerably above $e^\infty_\gs$. Our experience is that building up TMERA yields better performance than building up TTTN and then switching to TMERA. There are a few exceptions to this rule, e.g., for the BLBQ spin-1 chain at $\vartheta = \pi/4$. For the spin-1/2 XXZ chain, the scanning scheme gives somewhat better results than the built-up TMERA. Here, we started from the low-entangled ground state at $\Delta=2$ in the N\'{e}el phase and then scanned forth and back in the range $0\leq\Delta\leq 2$ until reaching convergence at $\Delta=1$.

\section{Computational time complexity}\label{sec:complexity}
\Emph{TMERA VQE.}~--- Assuming that gates which act on disjoint sets of qubits can be executed in parallel, the computation time for evaluating the TMERA expectation value of a local interaction term $\hh_i$ on the quantum computer is proportional to the total circuit depth $\O(tT)$. For the employed gradient-based optimization algorithm, we need to determine the gradient with respect to each of the $\O(qtT)$ Trotter gates (to a certain accuracy). The total quantum computation time for a single measurement of all gradients is hence $\O(q(tT)^2)$ \cite{Miao2021_08}. Of course, the quantum computer yields probabilistic digital outcomes rather than exact expectation values. Therefore, one has to consider the number of measurement samples $N_s$ needed in each optimization step. If we want to reach an accuracy
\begin{equation}\label{eq:accuracy}
	\epsilon :=e-e^\infty_\gs
\end{equation}
for the energy density, the required number of measurement samples scales as $N_s\sim 1/\epsilon^2$. Hence, the quantum cost for each TMERA optimization step is $\O(qt^2T^2/\epsilon^2)$. Quantum amplitude estimation (QAE) \cite{Knill2007-75,Wang2019-122} can reduce the required number of samples from $\O(1/\epsilon^2)$ to $\O(\log(1/\epsilon))$ while increasing the circuit depth by a factor $\O(1/\epsilon)$ \footnote{In the QAE scheme, it is not feasible to use mid-circuit qubit resets. Hence, the number of qubits required for TMERA is then not system-size independent but grows linearly in $T$, i.e., logarithmically in the (effective) system size. Still, only the causal-cone qubits need to reside in the quantum register and the others can be moved to a quantum memory \cite{Miao2021_08}.}. Hence, the resulting quantum computation time scales as
\begin{equation}\label{eq:VQEcost}
	\O\Big(\frac{qt^2T^2}{\epsilon}\,\log\frac{1}{\epsilon}\Big).
\end{equation}

\Emph{Classical (T)MERA algorithms.}~--- The computation time for EEG-based fMERA optimizations on classical computers is dominated by the costs for (exactly) contracting tensor networks that yield the energy gradient 
\begin{equation}\label{eq:Egrad}
	\partial_{U_{\tau,k}}\bra\Psi|\hH|\Psi\ket, 
\end{equation}
where $U_{\tau,k}$ denotes one of the MERA tensors in layer $\tau$. The computation time follows a power law $\O(\chi^r)$ with rather large exponents $r$ especially in $D\geq 2$ spatial dimensions. This holds regardless of whether the optimization is based on Riemannian gradients \cite{Hauru2021-10,Miao2021_08} or the traditional fixed-point-iteration SVD scheme \cite{Evenbly2009-79}. The costs for EEG-based TMERA optimizations depend additionally on the number of Trotter steps $t$ needed to achieve a certain energy accuracy \eqref{eq:accuracy}. In the VMC-based fMERA and TMERA methods, energy gradients \eqref{eq:Egrad} are not determined exactly but sampled stochastically, where, in analogy to the VQE scheme, one propagates pure causal-cone states in the MERA preparation direction and applies projective measurements to the sites that leave the causal cone in the layer transitions \cite{Ferris2012-85,Barthel2025-111}. Hence, the associated cost attains an additional factor $1/\epsilon^2$ corresponding to the required number of samples.
As will also be confirmed in the benchmark simulations, for critical models, the energy accuracy and optimal number of Trotter steps scale algebraically with respect to the bond dimension $\chi$,
\begin{equation}\label{eq:powerLaws}
	\epsilon\sim\chi^{-\beta}\quad\text{and}\quad
	t\sim \chi^p.
\end{equation}
In Ref.~\cite{Barthel2025-111}, we have determined optimal tensor contraction sequences for all steps in the four different classical algorithms using cost-pruned breadth-first constructive searches \cite{Lam1997-07,Hartono2005-155,Pfeifer2014-90} and depth-first searches. This was done for the six common types of MERA: the 1D binary MERA \cite{Vidal-2005-12}, 1D modified binary MERA \cite{Evenbly2013}, 1D ternary MERA \cite{Evenbly2009-79}, 2D quaternary MERA \cite{Cincio2008-100}, 2D two-step nonary MERA \cite{Evenbly2009-79}, and the 2D three-step nonary MERA \cite{Evenbly2009-102}; see Fig.~4 in Ref.~\cite{Barthel2025-111}. The rigorous results for the optimal cost scaling in the classical algorithms are summarized in Table~\ref{tab:MERAcost}.
\begin{figure}[t]
	\includegraphics[width=\columnwidth]{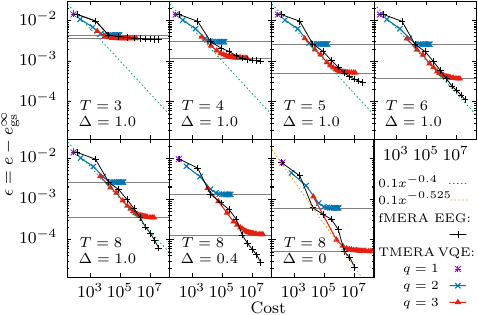}
	\caption{\label{fig:cost1} \textbf{Comparison of TMERA-VQE and fMERA-EEG time complexities for critical spin-1/2 XXZ chains}. The energy accuracy $\epsilon=e-e^\infty_\gs$ is shown as a function of the total computation time per iteration. The latter are $\O(\chi^7)$ for fMERA EEG simulations and $qt^2\log(\frac{1}{\epsilon})/\epsilon$ for the TMERA VQE. For $\Delta=1$ and different numbers of layers $T$, the top panels show saturation effects when varying $t$ for TMERA and $\chi$ for fMERA, where the two horizontal lines indicate the fMERA optima for $\chi=4$ ($q=2$) and $\chi=8$ ($q=3$), respectively. For the lower panels, $T=8$ is fixed and the anisotropy parameter $\Delta$ of the model \eqref{eq:XXZ} is varied. The dashed lines indicate the estimated accuracy-cost scaling of TMERA VQE for $\Delta=1$ and $\Delta=0$, respectively. We use homogeneous modified binary MERA and the scanning method. The data points show the minimum of the energy density over a few hundred randomly initialized optimizations. We use fMERA with $\chi\leq 12$, and TMERA with $t\leq 6$ for $q=2$, and $t\leq 16$ for $q=3$.}
\end{figure}

\begin{figure*}[p]
	\includegraphics[width=\textwidth]{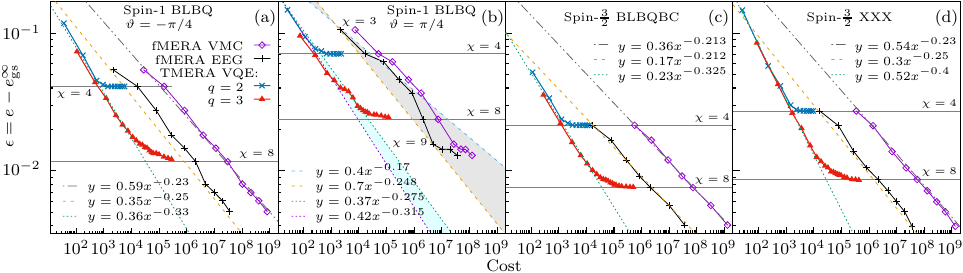}
	\caption{\label{fig:cost2} \textbf{Comparison of fMERA-EEG as well as fMERA-VMC to TMERA-VQE time complexities for critical spin-1 and 3/2 chains}. As in Fig.~\ref{fig:cost1}, the energy accuracy is shown as a function of the total computation time per iteration. We use homogeneous modified binary MERA with $T = 6$ layers, which is sufficient for the considered accuracy range and models \eqref{eq:BLBQ}, \eqref{eq:BLBQBC}, and \eqref{eq:XXX3d2}. The MERA are built-up. For the BLBQ model at $\vartheta = \pi/4$, the built-up TTN\,$\mapsto$MERA method and scanning are also used. Each data point represents the minimum of the energy density over a few hundred randomly initialized optimizations. We use $\chi\leq 12$ and, in TMERA, $t\leq 6$ for $q=2$, and $t\leq 16$ for $q=3$.}
\end{figure*}

\Emph{Numerical observations.}~--- 
For the critical spin models described in Sec.~\ref{sec:models}, TMERA-VQE and fMERA computation costs per iteration as a function of the energy accuracy $\epsilon$ [Eq.~\eqref{eq:accuracy}] are shown in Figs.~\ref{fig:cost1} and \ref{fig:cost2}.
\begin{table*}[p]
	\centering
	\setlength{\tabcolsep}{1.6ex}
	\renewcommand{\arraystretch}{1.8}
	\begin{tabular}{|l | c c | c c c |}
	\hline
	                                & \multicolumn{2}{|c|}{full MERA} & \multicolumn{3}{|c|}{Trotterized MERA}\\[-0.9em]
	\multicolumn{1}{|c|}{MERA type} & EEG & VMC                   & $p_\text{max}$  & EEG & VMC\\
	\hline
	1D binary          & $\O(\chi^9)$ & $\O(\chi^{6+2\beta})$ & 4   & $\O\Big(\chi^{8\lin{0}{1}9}\Big)$            & $\O(\chi^{4+p+2\beta})$\\
	1D mod.\ binary    & $\O(\chi^7)$ & $\O(\chi^{5+2\beta})$ & 4   & $\O\Big(\chi^{6\lin{0}{1}7\lin{3}{4}8}\Big)$ & $\O(\chi^{3+p+2\beta})$\\
	1D ternary         & $\O(\chi^8)$ & $\O(\chi^{5+2\beta})$ & 4   & $\O(\chi^{8})$                               & $\O(\chi^{4+p+2\beta})$\\
	2D            $2\times 2\mapsto 1$ & $\O(\chi^{26})$ & $\O(\chi^{16+2\beta})$ & 8  & $\O(\chi^{26})$                      & $\O\Big(\chi^{\big(14\lin{0}{2}16\big)+2\beta}\Big)$\\
	2D two-step   $3\times 3\mapsto 1$ & $\O(\chi^{16})$ & $\O(\chi^{15+2\beta})$ & 10 & $\O\Big(\chi^{16\lin{6}{10}20}\Big)$ & $\O\Big(\chi^{\big(14\lin{0}{1}15\lin{5}{10}20\big)+2\beta}\Big)$\\
	2D three-step $3\times 3\mapsto 1$ & $\O(\chi^{16})$ & $\O(\chi^{11+2\beta})$ & 8  & $\O(\chi^{16})$                      & $\O\Big(\chi^{\big(9\lin{0}{2}11\lin{4}{8}15\big)+2\beta}\Big)$\\[0.15em]
	\hline
	\end{tabular}
	\caption{\label{tab:MERAcost}\textbf{Computation costs for classical fMERA and TMERA algorithms based on EEG and VMC.} The stated scaling of the computation cost per optimization step is based on optimal tensor contraction sequences as discussed in Ref.~\cite{Barthel2025-111}. We assume critical models such that the energy accuracy $\epsilon$ and optimal number of Trotter steps $t$ in TMERA follow power laws \eqref{eq:powerLaws} with scaling exponents $\beta$ and $p$. The VMC importance sampling requires $\O(1/\epsilon^2)=\O(\chi^{2\beta})$ samples per optimization step, and the TMERA costs depend on $t=\chi^p$. As a function of $p$, the exponents in the cost scaling have constant and linearly increasing intervals. The notation ``$\alpha_1\lin{p_1}{p_2}\alpha_2$'' denotes a $p$ dependent function with value $\alpha_1$ until $p=p_1$ and a linear increase on the interval $(p_1,p_2)$, reaching the value $\alpha_2$ at $p=p_2$.
	The stated $p_\text{max}$ is an upper bound for $p$ due to the fact that an fMERA tensor that maps from $\nu$ input sites to $\mu$ output sites can always be implemented as a circuit of $\O(\chi^{\nu+\mu})$ single-qubit and CNOT gates \cite{Iten2016-93}.}
\end{table*}
\begin{table*}[p]
	\centering
	\setlength{\tabcolsep}{1.5ex}
	\begin{tabular}{l c c c c}
	\hline
	Model & energy exponent\ $\beta$ & Trotter exponent\ $p$ & In Fig.~\ref{fig:algCmp}b\\[0.2em]
	\hline
	Spin-1/2 XX chain			$\hH_{{1}/{2}}(\Delta=0)$		& 3.68 & 1.66		& green\\
	Spin-1/2 XXX chain			$\hH_{{1}/{2}}(\Delta=1)$		& 2.8  & 2.1		& red\\
	Spin-1 BLBQ chain at the TB point	$\hH_{1}(\vartheta=-{\pi}/{4})$	& 1.75 & 1.78		& blue\\
	Spin-1 BLBQ chain at the ULS point	$\hH_{1}(\vartheta=+{\pi}/{4})$	& 1.19\dots1.74 & 1.89	& purple\\
	Spin-3/2 BLBQBC chain at the TB point	$\hH_{{3}/{2}}$			& 1.48 & 1.54		& brown\\
	Spin-3/2 XXX chain			$\hH'_{{3}/{2}}$			& 1.75 & 1.31		& orange\\
	\hline
	\end{tabular}
	\caption{\label{tab:bp}\textbf{Energy and Trotter scaling exponents for the 1D modified binary MERA in critical spin chains.} For critical models \eqref{eq:XXZ}-\eqref{eq:XXX3d2}, the energy accuracy $\epsilon$ and optimal number of Trotter steps $t$ follow power laws \eqref{eq:powerLaws} in terms of the bond dimension $\chi$. For the 1D modified binary MERA, the corresponding scaling exponents $\beta$ and $p$ are determined from the data in Figs.~\ref{fig:cost1} and \ref{fig:cost2}.}
\end{table*}
When increasing the number of Trotter steps $t$ while $q$ is fixed, the TMERA accuracy $\epsilon$ improves until reaching the fMERA accuracy for bond dimension $\chi=2^q$. It is preferable to avoid this saturation regime and, instead, increase $\chi$ (the number of qubits $q=\log_2\chi$ per renormalized site) before continuing to increase $t$ \cite{Miao2021_08}. The fits to the numerical results in Figs.~\ref{fig:cost1} and \ref{fig:cost2} confirm the power-law scaling \eqref{eq:powerLaws} for $\epsilon$ and the optimal $t$ as functions of $\chi$.

The attainable accuracy $\epsilon$ also depends on the number of layers $T$. For the considered critical models and fixed $T$, the accuracy saturates as a function of $\chi$. This is due to the infinite correlation length in these systems, where the long-range correlations of the targeted ground states can only be encoded in higher-layer tensors. The top panels in Fig.~\ref{fig:cost1} clearly demonstrate this saturation for TMERA and fMERA. For the spin-1/2 XXZ model \eqref{eq:XXZ} at $\Delta = 1$, the numerical results suggest that $T=6$ layers are sufficient for $\chi\leq8$ and $T=8$ for $\chi\leq12$. For smaller $\Delta\in[0,1)$, one needs to increase the number of layers further to avoid the saturation. This can be attributed to the fact that the decay of the transverse correlation function $|\langle\hS^{x}_i \hS^{x}_{j}\rangle|\sim|i-j|^{-\eta}$ is slower for smaller $\Delta$, where the critical exponent is given by
\begin{equation}\label{eq:XXZeta}
	\eta = 1-\arccos(\Delta)/\pi\ \in \ [1/2,1).
\end{equation}
The longitudinal correlations always decay faster with $|\langle\hS^{z}_i \hS^{z}_{j}\rangle|\sim|i-j|^{-1/\eta}$ \cite{Luther1974-12,Mikeska2004}.
For the higher-spin models, we find that $T=6$ layers are sufficient when $\chi\leq 12$; see Fig.~\ref{fig:cost2}.

\section{Quantum advantage}\label{sec:qAdvantage}
\Emph{Scope.}~--- Let us now assess whether the TMERA VQE can provide a quantum advantage over the relevant classical MERA algorithms. Until now, almost all classical simulations optimized fMERA based on EEG \cite{Evenbly2009-79}. The only exception that we are aware of is Ref.~\cite{Ferris2012-85}, where Ferris and Vidal tested fMERA VMC for 1D systems but did not attempt a thorough comparison of computation costs.
Beyond outsourcing the (classically expensive) tensor contractions to a quantum processor, TMERA VQE differs in two important aspects from fMERA EEG:
\begin{enumerate}[label=(\alph*)]
 \item the MERA tensors are constrained to have Trotter circuit structure (Fig.~\ref{fig:MERA}), and
 \item one samples pure causal-cone states according to the Born rule for quantum measurements, very similar to VMC \cite{Miao2021_08,Ferris2012-85,Barthel2025-111}.
\end{enumerate}
To be able to identify the source of advantages of TMERA VQE, we compare it to the four relevant classical MERA algorithms which, in the spirit of quantum-inspired classical algorithms, can entail tensor Trotterization and/or pure-state importance sampling, i.e., fMERA EEG, fMERA VMC, TMERA EEG, and TMERA VMC \cite{Barthel2025-111}.

For a sufficiently large number of layers $T$, the achievable energy accuracy $\epsilon$ is a function of the bond dimension $\chi$ and the number of Trotter steps $t$ in TMERA. The scaling laws \eqref{eq:powerLaws} allow us to express the computation costs for all methods as functions of $\chi=\chi(\epsilon)$ only. For the classical algorithms, this is done in Table~\ref{tab:MERAcost} and, up to logarithmic corrections, the TMERA-VQE cost \eqref{eq:VQEcost} becomes
\begin{equation}\label{eq:VQEcost2}
	\O(\chi^{2p+\beta}).
\end{equation}
For the comparison, we focus on the leading-term exponent ($2p+\beta$ for the TMERA VQE), i.e., disregard subleading terms and coefficients. The latter depend on implementation details and may of course be large. Hence, our statements about computational advantages concern, at this stage, the large-$\chi$ (high-accuracy) regime. The classical and quantum computation times, respectively, depend linearly and quadratically on the number of MERA layers $T$. However, $T$ only grows \emph{logarithmically} in the total system size $N$, and, even for critical systems, finite-size corrections to the energy density $e$ decrease algebraically with increasing $N$. Hence, the $T$ dependence of the computation times is a logarithmic correction that we can neglect.

\Emph{Numerical results for critical 1D systems.}~--- Classical MERA simulations of 1D systems are sufficiently efficient such that we can reliably determine the energy and Trotter exponents $\beta$ and $p$ in the scaling laws \eqref{eq:powerLaws}.
For critical spin-1/2 XXZ chains \eqref{eq:XXZ} with $\Delta=1,0.4$, and 0, the achievable accuracies are shown as functions of the TMERA VQE and classical fMERA costs in the lower panels of Fig.~\ref{fig:cost1}. For this model, TMERA VQE and fMERA EEG display a rather similar scaling with the performance of both improving with decreasing $\Delta$. Similarly, Fig.~\ref{fig:cost2} shows the scaling analysis for the critical spin-1 and spin-3/2 systems \eqref{eq:BLBQ}-\eqref{eq:XXX3d2}, suggesting a polynomial quantum advantage of the TMERA VQE over the classical fMERA EEG and VMC methods for all considered models.
The spin-3/2 XXX chain \eqref{eq:XXX3d2} has the same central charge as its spin-1/2 counterpart, and the TMERA VQE accuracy has almost the same scaling for both models. Interestingly, the fMERA shows a considerably worse scaling for the spin-3/2 case.

From these simulations, we extract the energy and Trotter scaling exponents $\beta$ and $p$ in Eq.~\eqref{eq:powerLaws}. The results are stated in Table~\ref{tab:bp}.
\begin{figure*}[t]
        \setlength{\tabcolsep}{2ex}
	\begin{tabular}{l l l}
	{\footnotesize (a) \ 1D binary MERA}&
	{\footnotesize (b) \ 1D modified binary MERA}&
	{\footnotesize (c) \ 1D ternary MERA}\\[0.2em]
	\includegraphics[width=0.3055\textwidth]{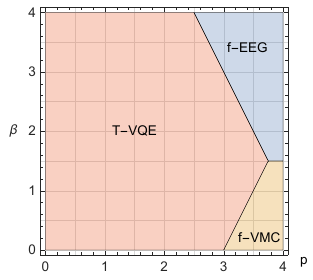}&
	\includegraphics[width=0.3055\textwidth]{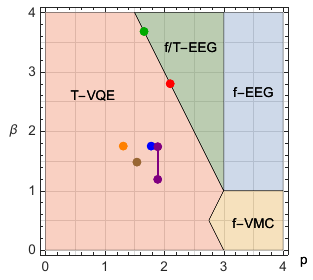}&
	\includegraphics[width=0.3055\textwidth]{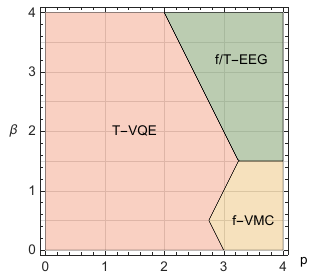}\\[0.6em]
	{\footnotesize (d) \ 2D $2\times 2\mapsto 1$ MERA}&
	{\footnotesize (e) \ 2D two-step   $3\times 3\mapsto 1$ MERA}&
	{\footnotesize (f) \ 2D three-step $3\times 3\mapsto 1$ MERA}\\[0.2em]
	\includegraphics[width=0.3055\textwidth]{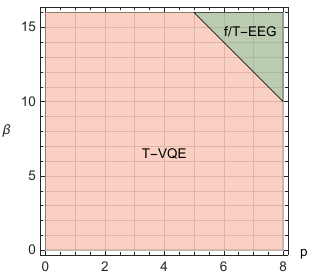}&
	\includegraphics[width=0.3055\textwidth]{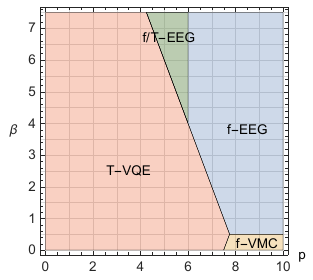}&
	\includegraphics[width=0.3055\textwidth]{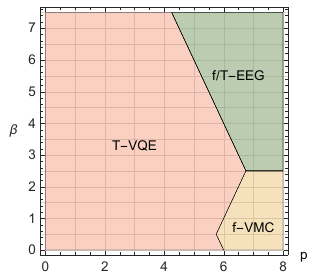}\\
	\end{tabular}
	\caption{\label{fig:algCmp}\textbf{Algorithmic phase diagrams.} For critical models, the energy accuracy $\epsilon$ [Eq.~\eqref{eq:accuracy}] and optimal number of Trotter steps $t$ follow power laws in terms of the bond dimension $\chi$ as stated in Eq.~\eqref{eq:powerLaws}. In dependence of the corresponding model-specific scaling exponents $\beta$ and $0\leq p\leq p_{\max}$, the quantum and classical computation costs lead to the shown algorithmic phase diagrams. They specify which of the five methods in question (TMERA VQE, fMERA EEG, fMERA VMC, TMERA EEG, and TMERA VMC) has the smallest scaling exponent for the computation costs per optimization step. The colored points in panel (b) indicate the scaling exponents of different critical spin models as specified in Table~\ref{tab:bp}.}
\end{figure*}

\Emph{Algorithmic phase diagrams.}~--- 
Based on the scaling of computation costs in Eq.~\eqref{eq:VQEcost2} and Table~\ref{tab:MERAcost}, Fig.~\ref{fig:algCmp} provides algorithmic phase diagrams. As a function of the scaling exponents $\beta$ and $p\leq p_\text{max}$ for the energy accuracy and number of Trotter steps, the diagrams show which of the five considered methods has the lowest computation costs per optimization step in the large-$\chi$ regime. For all six types of MERA structures (specified in Sec.~\ref{sec:complexity}), the TMERA VQE dominates the realistic $(\beta,p)$ ranges. For the 1D modified binary MERA (Fig.~\ref{fig:algCmp}b), in particular, all critical groundstate problems from Sec.~\ref{sec:models} lie in the region where TMERA VQE is most efficient. The spin-1/2 XXZ chain in the critical XY phase ($|\Delta|\leq 1$) happens to fall directly onto the transition line from the TMERA-VQE region to a region where fMERA and TMERA EEG have the same and best cost scaling as already suggested by the results in Fig.~\ref{fig:cost1}. Generally, the classical algorithms only prevail in relatively small or unrealistic $(\beta,p)$ regions with fMERA VMC at large $p$ and small $\beta$ as well as fMERA and/or TMERA EEG at large $p$ and large $\beta$. Note for example that, in the diagram for the 2D quaternary MERA \cite{Cincio2008-100}, no classical algorithm can compete with the VQE for $0\leq\beta\leq 10$ (see Fig.~\ref{fig:algCmp}d).
Lastly, according to this analysis, the classical TMERA VMC method \cite{Barthel2025-111} is not competitive for any of the considered 1D and 2D MERA.

We can conclude that the observed polynomial advantage of TMERA VQE over the classical algorithms is predominantly due to the replacement of the costly classical tensor contractions with the execution of Trotter quantum circuits on the quantum computer. Neither the tensor Trotterization (TMERA) nor importance sampling (VMC), when applied separately or in combination on classical computers, can outperform the VQE for realistic values of the energy and Trotter scaling exponents. The separation is more pronounced for systems in $D\geq 2$ dimensions due to the rapid increase of MERA contraction costs for higher-dimensional systems. In contrast to the classical EEG methods, quantum computers produce probabilistic measurement outcomes. However, the results show that drawbacks of the resulting necessity to sample energy gradients are outweighed by the avoidance of high classical tensor contraction costs.

\newpage
\section{Reducing rotation angle amplitudes}\label{sec:anglePenalty}
\begin{figure}[t]
	\includegraphics[width=\columnwidth]{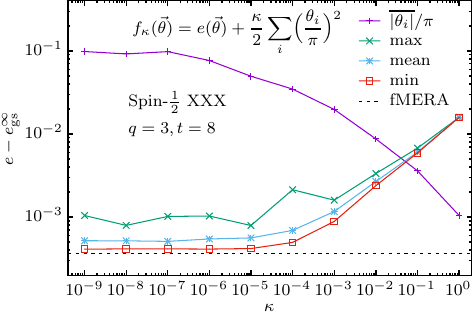}
	\caption{\label{fig:penalty}\textbf{Effect of large-angle penalties}. The plot shows results for the spin-1/2 XXX model using homogeneous modified binary TMERA with $T = 6$, $t = 8$, and $q = 3$. The modified objective function \eqref{eq:anglePenalty} comprises the energy density and a term penalizing large rotation angles.
	The three lower curves show the maximum, mean value, and minimum of the energy density accuracy for a few hundred randomly initialized TMERA optimizations. The displayed average two-qubit rotation angle is obtained from the converged lowest-energy TMERA.
	}
\end{figure}
For experimental realizations, the TMERA Trotter gates are ultimately parametrized by rotation angles $\theta$, e.g., through the canonical decomposition for an arbitrary SU(4) matrix into single-qubit rotations like
\begin{equation}
	\hR_{\hat{\vec{\sigma}}}(\alpha,\beta,\gamma)=e^{-\mri\alpha \hs^z/2}e^{-\mri\beta \hs^y/2}e^{-\mri\gamma \hs^z/2}
\end{equation}
and two-qubit rotations
\begin{equation}
	\hR_{\hs^x\otimes\hs^x}(\theta)=e^{-\mri\theta \hs^x\otimes\hs^x/2},
\end{equation}
$\hR_{\hs^y\otimes\hs^y}(\theta)$, and $\hR_{\hs^z\otimes\hs^z}(\theta)$; see Fig.~\ref{fig:MERA}d and Refs.~\cite{Kraus2001-63,Zhang2003-67}.
With each gate corresponding to a unitary time evolution, the amplitudes of these angles determine the experimental simulation times or the intensities of manipulation pulses. Small rotation angles are desirable since both long evolution times and high pulse intensities decrease the quantum gate fidelity. We propose to reduce average rotation angle amplitudes by adding an angle penalty term to the energy functional. Figure~\ref{fig:penalty} shows optimization results for the modified objective function
\begin{equation}\label{eq:anglePenalty}
	f_\kappa(\vec{\theta}) = e(\vec{\theta}) + \frac{\kappa}{2}\sum_i (\theta_i/\pi)^2,
\end{equation}
where the sum in the penalty term runs over the angles $\{\theta_i\}$ of all two-qubit rotations. As expected, the average angle amplitude $\overline{|\theta_i|}$ decreases as the penalty parameter $\kappa$ increases. We find that the average angle can be reduced by a factor of around 2 with negligible effect on the energy accuracy.

For the data in Fig.~\ref{fig:penalty}, we used built-up TMERA and Euclidean L-BFGS, operating directly on the angles $\vec{\theta}$, instead of Riemannian L-BFGS. Alternatively, one could change the large-angle penalty term in Eq.~\eqref{eq:anglePenalty} to $\frac{\kappa}{2}\sum_j \|\hU_j-\id\|^2$ with the sum running over all Trotter gates $\hU_j$ of the TMERA and, then, employ the Riemannian optimization as described in Ref.~\cite{Miao2021_08}.

\section{Random circuit TMERA tensors}\label{sec:PRPC}
\begin{figure}[t]
	\includegraphics[width=\columnwidth]{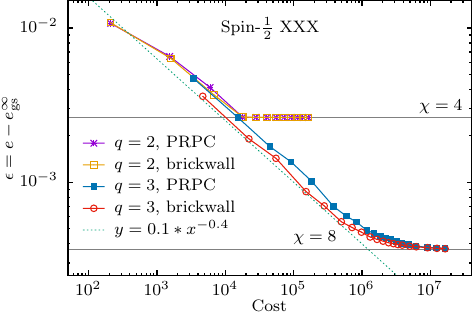}
	\caption{\label{fig:PRPC} \textbf{Energy accuracies for TMERA with brick-wall and PRPC tensors}. The plot shows the energy accuracy for the spin-1/2 Heisenberg antiferromagnet using homogeneous modified binary TMERA with two different tensor circuit structures -- the brickwall circuits and parallel random-pair circuits illustrated in Fig.~\ref{fig:MERA}. The number of MERA layers is $T = 6$, and the number of Trotter steps is $t\leq 6$ for $q=2$ and $t\leq 16$ for $q=3$. For each data point, we chose the minimum energy from a few hundred randomly initialized optimizations.}
\end{figure}

So far, we chose the TMERA tensors as brick-wall circuits with Trotter gates acting on nearest-neighbor qubits. The expressiveness and efficiency of the ansatz might be improved for alternative choices. In ion-trap systems one could, for example, use the all-to-all connectivity \cite{Wright2019-10,Linke2017-114} and work, instead, with parallel random-pair circuits (PRPC) as illustrated in Fig.~\ref{fig:MERA}c. Each Trotter step consists again of two Trotter-gate coverings. In each covering, $n/2$ two-qubit gates act on disjoint qubit pairs, randomly chosen from the $n$-qubit support of the TMERA tensor. For the spin-1/2 Heisenberg antiferromagnet, Fig.~\ref{fig:PRPC} shows a comparison of TMERA accuracies for brick-wall circuit tensors and PRPC tensors. Even though a PRPC contains long-range two-qubit gates and hence allows for faster scrambling, the scaling of the attained accuracies is very similar, at least for this model and relatively small bond dimensions.

\section{Discussion}\label{sec:discuss}
\Emph{Quantum advantage.}~--- 
For the groundstate problems of various critical 1D spin systems, we have numerically confirmed in Sec.~\ref{sec:complexity} that the TMERA VQE offers a polynomial quantum advantage over classical fMERA and TMERA simulations based on EEG and VMC. Furthermore, the classical simulation costs increase drastically when going from 1D to higher-dimensional systems, e.g., from $\O(\chi^{7\dots 9})$ for 1D fMERA EEG to $\O(\chi^{16\dots 26})$ for 2D fMERA EEG (see Table~\ref{tab:MERAcost}). This is due to the high costs associated with the classical contraction of high-order tensors. As these are not relevant for the time complexity of the TMERA VQE, the algorithmic phase diagrams in Sec.~\ref{sec:qAdvantage} (Fig.~\ref{fig:algCmp}), spanned by the energy-accuracy and Trotter-step scaling exponents for critical systems [Eq.~\eqref{eq:powerLaws}], suggest an even more substantial quantum advantage for 2D systems. Unfortunately, it is currently very challenging to properly determine the scaling of MERA costs and accuracies in classical simulations for 2D systems; the best option might be VMC for the 2D three-step $3\times 3\mapsto 1$ fMERA \cite{Barthel2025-111}.

Of course, substantiating a quantum advantage for TMERA VQE over the considered classical MERA algorithms does not exclude other more efficient classical algorithms. In fact, it is conceivable that matrix product state (MPS) \cite{Baxter1968-9,Fannes1992-144,White1992-11,Rommer1997,Schollwoeck2011-326} are more efficient for many 1D systems -- at least for sufficiently small bond dimensions $\chi$. However, 1D MERA can encode critical correlations while MPS necessarily show an exponential decay beyond a $\chi$-dependent length scale. Also, for an $L_x\times L_y$ 2D system, MPS simulation costs increase exponentially in $\min(L_x,L_y)$ and are hence not competitive. For unfrustrated 2D spin and bosonic systems, quantum Monte Carlo (QMC) \cite{Suzuki1977-58,Sandvik1991-43,Prokofev1996-64,Syljuasen2002-66,Alet2005-71} may be more efficient. Note however that the sign problem \cite{Takasu1986-75,Loh1990-41,Chandrasekharan1999-83,Troyer2005} renders QMC inefficient for frustrated quantum magnets and fermionic systems in $D\geq 2$ dimensions, which are arguably of highest scientific and technological interest as they comprise the topologically ordered systems and high-temperature superconducting materials. Another prominent competitor are planar tensor networks called projected entangled pair states (PEPS) \cite{Niggemann1997-104,Nishino2000-575,Martin-Delgado2001-64,Verstraete2004-7,Verstraete2006-96}.

A thorough comparison of TMERA VQE with these other classical simulation methods is challenging and falls outside the scope of this work. In the comparison of the different quantum and classical MERA methods, we only needed to assess the costs per optimization step as they basically have the same convergence behavior. In contrast, the convergence properties of alternative methods can be quite different which results in the challenge of having to compare entire optimization trajectories which depend on various details. For example, the MERA layer transfer maps are generally gapped but, for critical 1D systems, the gap of the MPS site transfer maps will generally vanish with increasing bond dimension. For $D\geq 2$ dimensions, MERA actually form a subclass of PEPS \cite{Barthel2010-105}. However, while MERA allows for an exact evaluation of observables, this is generally not possible for PEPS. Hence, PEPS optimizations already involve approximations in the evaluation of energy gradients which leads to a variety of different PEPS schemes and nontrivial convergence properties. We hope that future research can elucidate the different cost scalings of these methods beyond anecdotal comparisons for specific models.

\Emph{Improvements and barren plateaus.}~--- 
We found in Sec.~\ref{sec:converge} that the convergence of the TMERA VQE can be improved substantially by gradually increasing the number of variational parameters and scanning in the model parameter space. One could explore further optimization methods like basin hopping \cite{Wales1997-101} and quantum natural gradient \cite{Stokes2020-4,Wierichs2020-2}.

The results in Sec.~\ref{sec:anglePenalty} show that adding a simple angle-penalty term to the energy functional can considerably reduce the average rotation angles that parametrize the Trotter gates (reduce the distance of the Trotter gates from the identity) without substantially worsening the energy accuracy. We hope that continued research along these lines can further reduce the experimental resources needed for the TMERA VQE.

The two different types of Trotter circuit structures for the TMERA tensors, compared in Sec.~\ref{sec:PRPC}, produced very similar results. As they were both based on two-qubit gates, it would be interesting to explore whether multi-qubit gates, naturally available on different quantum computing platforms, can improve the efficiency of the ansatz.

The success of the (T)MERA optimizations for large quantum systems with substantial groundstate entanglement indicates that the approach is not hampered by vanishing-gradient issues. In fact, Refs.~\cite{Barthel2023_03,Miao2024-109} rigorously prove the absence of barren plateaus \cite{McClean2018-9,Cerezo2021-12} for isometric tensor network states, including matrix product states (MPS), tree tensor networks (TTN), MERA and TMERA.

\Emph{Conclusion.}~---
Given our shortage of sufficiently efficient classical methods for the solution of groundstate problems for strongly-correlated quantum many-body systems in $D\geq 2$ spatial dimensions, including systems with topological order and high-temperature superconducting materials, and given a shortage of useful applications for NISQ devices, the TMERA VQE represents an exciting new route:
\begin{enumerate}[label=(\alph*)]
 \item Due to its narrow causal cone, the algorithm can be implemented on intermediate-scale quantum devices and still describe large systems -- even infinite systems when exploiting translation invariance. The number of required qubits is system-size independent and increases only to a logarithmic scaling when using quantum amplitude estimation to speed up gradient evaluations \cite{Miao2021_08}.
 \item Arguments by Kim and Swingle \cite{Kim2017_11} also apply to TMERA and suggest their resilience against noise. This is due to the fact that MERA layer transition channels are generally gapped such that noise does not proliferate but gets damped when transitioning to the next layer.
 \item We have rigorously shown that MERA are not hampered by barren plateaus and that average energy-gradient amplitudes only decay polynomially with increasing bond dimension \cite{Barthel2023_03,Miao2024-109}.
 \item Sections~\ref{sec:complexity} and \ref{sec:qAdvantage} substantiate a polynomial quantum advantage over the classical MERA methods.
 \item The resources required for an experimental implementation are relatively low and realistically achievable.
\end{enumerate}
As an example concerning the latter point, a 1D binary TMERA VQE with bond dimension $\chi=8$ ($q=3$), $T=6$ layers, and $t=2$ Trotter steps requires only 12 qubits and circuit depth 120 when optimizing for low qubit number, or 18 qubits and circuit depth 48 when optimizing for low depth.
Also, a 2D $2\times 2\mapsto 1$ TMERA VQE with bond dimension $\chi=4$ ($q=2$), $T=6$ layers, and $t=2$ Trotter steps requires only 28 qubits and circuit depth 624 when optimizing for low qubit number, or 72 qubits and circuit depth 96 when optimizing for low depth.
First experimental implementations of MERA on ion-trap quantum computers have demonstrated critical groundstate correlation functions \cite{Haghshenas2023_05} as well as a quantum phase transition and the associated change from area-law to log-area-law entanglement scaling \cite{Miao2024_12}.

As discussed in Secs.~\ref{sec:converge}, \ref{sec:anglePenalty}, and \ref{sec:PRPC}, there is ample room for further improvements to reduce experimental resources and achieve better groundstate accuracy.

\begin{acknowledgments}
We gratefully acknowledge discussions with Marko Cetina, Kenneth R.\ Brown, Jutho Haegeman, and Yikang Zhang, helpful feedback by the anonymous referees,
as well as support through US Department of Energy grant DE-SC0019449 and the US National Science Foundation (NSF) Quantum Leap Challenge Institute for Robust Quantum Simulation (award no.\ OMA-2120757).
\end{acknowledgments}

\end{document}